\documentclass[showpacs,preprintnumbers,amsmath,amssymb,fleqn]{revtex4}
\usepackage{graphics}
\normalsize

\begin{document}
\title{Controlling entanglement sudden death and birth in cavity QED}
\author{Jian-Song Zhang} \author{Jing-Bo Xu}
 \email{xujb@zju.edu.cn}
\affiliation{Zhejiang Institute of Modern Physics and Physics
Department,\\ Zhejiang University, Hangzhou 310027, China}

\date{\today}

\begin{abstract}
We present a scheme to control the entanglement sudden birth and
death in cavity quantum electrodynamics system, which consists of
two noninteracting atoms each locally interacting with its own
vacuum field, by applying and adjusting classical driving fields.
\end{abstract}

 \maketitle

\section{INTRODUCTION}
In recent years, entanglement has been considered as a key resource
of quantum information processing \cite{1,2,3,4}. A cavity quantum
electrodynamics(QED)  system is a useful tool to create the
entanglement between atoms in cavities and establish quantum
communications between different optical cavities. Recently, the
manipulation of quantum entanglement for the system of cavity QED
has been extensively investigated\cite{5,6,7,8,9,10,11,12,13}.

Many efforts have been devoted to the study of the evolution of the
entanglement under the influence of the environment
\cite{14,15,16,Yu2009,17,18,19}. It is pointed out by Yu and Eberly
\cite{14} that the entanglement of an entangled two-qubit
interacting with uncorrelated reservoirs may disappear within a
finite time during the dynamics evolution. This phenomenon, called
entanglement sudden death (ESD) has been observed in experiment
\cite{20,21}. Recently, the entanglement sudden birth(ESB) in cavity
QED has been discussed  by Y\"{o}nac, Yu, and Eberly
\cite{Yonac2006, Yonac2007}. More recently, Lopez \emph{et al}.
\cite{22} have studied the entanglement dynamics of a quantum system
consisting of two cavities interacting with two independent
reservoirs and shown that ESD in a bipartite system independently
coupled to reservoirs is related to the ESB. It has been pointed out
that the cavity coherent state can be used to control the ESB and
ESD in cavity QED\cite{Yonac2008}.

In the present paper, we propose a scheme to control ESB and ESD of
a quantum system consisting of two noninteracting atoms each locally
interacting with its own vacuum field. The two atoms, which are
initially prepared in entangled states, are driven by two classical
fields additionally. It is shown that ESB and ESD phenomenon may
appear in this system and the time of ESB and ESD can be controlled
by classical driving fields. In addition, the amount of the
entanglement of the two atoms or cavities can be significantly
increased by applying classical fields.

\section{Effective Hamiltonian}
Now, we consider a system consisting of a two-level atom inside a
single mode cavity. The atom is driven by a classical field
additionally. The Hamiltonian of the system can be described by
\cite{12}
\begin{eqnarray}
H&=&\omega
a^{\dag}a+\frac{\omega_0}{2}\sigma_z+g(\sigma_+a+\sigma_-a^{\dag})\nonumber\\
&&+\lambda(e^{-i\omega_ct}\sigma_++e^{i\omega_ct}\sigma_-),
\end{eqnarray}
where $\omega$, $\omega_0$ and $\omega_c$ are the frequency of the
cavity, atom and classical field, respectively. The operators
$\sigma_z$ and
 $\sigma_{\pm}$ are defined by  $\sigma_{z}=|e\rangle\langle e|-|g\rangle\langle
 g|$, $\sigma_+=|e\rangle\langle g|$, and $\sigma_-=\sigma_+^{\dag}$ where
$|e\rangle$ and $|g\rangle$ are the excited and ground states of the
atom. Here, $a$ and $a^{\dag}$ are the annihilation and creation
operators of the cavity; g and $\lambda$ are the coupling constants
of the interactions of the atom with the cavity and with the
classical driving field, respectively. Note that we have set
$\hbar=1$ throughout this paper.

In the rotating reference frame the Hamiltonian of the system is
transformed to the Hamiltonian $H_1$ under a unitary transformation
$U_1= \exp{(-i\omega_c t\sigma_z/2)}$
\begin{eqnarray}
H_1&=&U_1^{\dag}HU_1-iU_1^{\dag}\frac{\partial U_1}{\partial
t}\nonumber\\
&=&H_1^{(1)}+H_1^{(2)},
\end{eqnarray}
with
\begin{eqnarray}
H_1^{(1)}&=&\omega
a^{\dag}a+g(e^{i\omega_ct}\sigma_+a+e^{-i\omega_ct}\sigma_-a^{\dag}),\nonumber\\
H_1^{(2)}&=&\frac{\Delta_1}{2}\sigma_z+\lambda(\sigma_++\sigma_-),
\end{eqnarray}
and $\Delta_1=\omega_0-\omega_c$. Using the method similar to that
used in Ref.\cite{23}, diagonalizing the Hamiltonian $H_1^{(2)}$,
and neglecting the terms which do not conserve energies (rotating
wave approximation), we can recast the Hamiltonian $H_1$ as follows:
\begin{eqnarray}
H_1&=&\omega
a^{\dag}a+\frac{\Omega_1\sin{\theta}}{2}(\sigma_++\sigma_-)
+g\cos^2{\frac{\theta}{2}}[e^{i\omega_ct}\nonumber\\
&&\times(-\frac{\sin{\theta}}{2}\sigma_z+\cos^2{\frac{\theta}{2}}\sigma_+-\sin^2{\frac{\theta}{2}}\sigma_-)a
+h.c],
\end{eqnarray}
with $\theta=\arctan{(\frac{2\lambda}{\Delta_1})}$. Here $h.c$
stands for Hermitian conjugation.

The Hamiltonian $H_1$ can be diagonalized by a final unitary
transformation $U_2$ with
$U_2=\exp{[\frac{i\omega_ct}{2}(\sigma_++\sigma_-)]}$. Then, we can
rewrite the Hamiltonian of the system
\begin{eqnarray}
H_2&=&\omega
a^{\dag}a+\frac{\omega'\sin{\theta}}{2}(\sigma_++\sigma_-)+g'[(-\frac{\sin{\theta}}{2}\sigma_z\nonumber\\
&&+\cos^2{\frac{\theta}{2}}\sigma_+
-\sin^2{\frac{\theta}{2}}\sigma_-)a +h.c],
\end{eqnarray}
where $\omega'=\sqrt{\Delta_1^2+4\lambda^2}+\omega_c$ and $g'=g
\cos^2{\frac{\theta}{2}}$. It is worth noting that the unitary
transformations $U_1$ and $U_2$ are both local unitary
transformations. As we known the entanglement of a quantum system
does not change under local unitary transformations \cite{24}. Thus,
the entanglement of the system considered here will not be changed
by applying the local unitary transformations $U_1$ and $U_2$.

\section{Controlling entanglement sudden death and birth}
In this section, we investigate ESD and ESB of a quantum system
consisting of two noninteracting atoms each locally interacting with
its own vacuum field. Each atom interacts with its own vacuum field
where the interaction of the system is described by $H_2$. We show
how to control entanglement sudden death and birth of a quantum
system formed by two two-level atoms and two cavities via classical
driving fields. Assume the two-level atoms are prepared in entangled
states and the cavities are prepared in vacuum states, i.e., the
whole system is initially prepared in the state
\begin{eqnarray}
|\psi(0)\rangle=(\alpha|-_{a_1}\rangle|-_{a_2}\rangle+\beta|+_{a_1}\rangle|+_{a_2}\rangle)|0_{c_1}\rangle|0_{c_2}\rangle,
\end{eqnarray}
where the subscripts $a_1$, $a_2$, $c_1$, and $c_2$ refer to atom 1,
atom 2, cavity 1, and cavity 2, respectively. Here, $|\pm\rangle$
can be interpreted as the dressed states of the two-level atom. They
are defined as follows:
\begin{eqnarray}
|+\rangle&=&\cos{\frac{\theta}{2}}|e\rangle+\sin{\frac{\theta}{2}}|g\rangle,\nonumber\\
|-\rangle&=&-\sin{\frac{\theta}{2}}|e\rangle+\cos{\frac{\theta}{2}}|g\rangle.
\end{eqnarray}

After some algebra, we find the state of the whole system at time t
is

\begin{eqnarray}
|\psi(t)\rangle&=&\alpha|-_{a_1}\rangle|-_{a_2}\rangle|0_{c_1}\rangle|0_{c_2}\rangle\nonumber\\
&&+\beta
f^2_1(t)|+_{a_1}\rangle|+_{a_2}\rangle|0_{c_1}\rangle|0_{c_2}\rangle\nonumber\\
&&+\beta
f^2_2(t)|-_{a_1}\rangle|-_{a_2}\rangle|1_{c_1}\rangle|1_{c_2}\rangle\nonumber\\
&&+\beta f_1(t)
f_2(t)(|+_{a_1}\rangle|-_{a_2}\rangle|0_{c_1}\rangle|1_{c_2}\rangle\nonumber\\
&&+|-_{a_1}\rangle|+_{a_2}\rangle|1_{c_1}\rangle|0_{c_2}\rangle),
\end{eqnarray}
with
\begin{eqnarray}
f_1(t)&=&e^{i\Delta_2t/2}[\cos{(\Omega
t)}-\frac{i\Delta_2}{2\Omega}\sin{(\Omega t)}], \nonumber\\
f_2(t)&=&-ig\cos^2{\frac{\theta}{2}}e^{-i\Delta_2t/2}\sin{(\Omega t)}/\Omega,\nonumber\\
\Delta_2&=&\sqrt{(\omega_0-\omega_c)^2+4\lambda^2}+\omega_c-\omega, \nonumber\\
\Omega&=&\sqrt{\frac{\Delta_2^2}{4}+(g\cos^2{\frac{\theta}{2}})^2}.
\end{eqnarray}
Tracing over the degrees of the freedom of cavities, we obtain the
reduced density matrix of two atoms
\begin{eqnarray}
\rho_{a_1a_2}(t)&=&[|\alpha|^2+|\beta
f^2_2(t)|^2]|-_{a_1}\rangle|-_{a_2}\rangle\langle-_{a_1}|\langle-_{a_2}|\nonumber\\
&&+|\beta
f^2_1(t)|^2|+_{a_1}\rangle|+_{a_2}\rangle\langle+_{a_1}|\langle+_{a_2}|\nonumber\\
&&+|\beta f_1(t)
f_2(t)|^2(|+_{a_1}\rangle|-_{a_2}\rangle\langle+_{a_1}|\langle-_{a_2}|\nonumber\\
&&+|-_{a_1}\rangle|+_{a_2}\rangle\langle-_{a_1}|\langle+_{a_2})\nonumber\\
&&+[\alpha \beta^*
f_1^{*2}(t)|-_{a_1}\rangle|-_{a_2}\rangle\langle+_{a_1}|\langle+_{a_2}|+h.c].
\end{eqnarray}
Similarly, the reduced density matrix of two cavities is
\begin{eqnarray}
\rho_{c_1c_2}(t)&=&[|\alpha|^2+|\beta
f^2_1(t)|^2]|-_{a_1}\rangle|-_{a_2}\rangle\langle-_{a_1}|\langle-_{a_2}|\nonumber\\
&&+|\beta
f^2_2(t)|^2|+_{a_1}\rangle|+_{a_2}\rangle\langle+_{a_1}|\langle+_{a_2}|\nonumber\\
&&+|\beta f_1(t)
f_2(t)|^2(|+_{a_1}\rangle|-_{a_2}\rangle\langle+_{a_1}|\langle-_{a_2}|\nonumber\\
&&+|-_{a_1}\rangle|+_{a_2}\rangle\langle-_{a_1}|\langle+_{a_2})\nonumber\\
&&+[\alpha \beta^*
f_2^{*2}(t)|-_{a_1}\rangle|-_{a_2}\rangle\langle+_{a_1}|\langle+_{a_2}|+h.c].
\end{eqnarray}

In order to study the entanglement of above system described by
density matrix $\rho$, we adopt the measure concurrence which is
defined as \cite{25}
\begin{equation}
C=\max{\{0, \lambda_1-\lambda_2-\lambda_3-\lambda_4\}},
\end{equation}
where the $\lambda_i$(i=1,2,3,4) are the square roots of the
eigenvalues in decreasing order of the magnitude of the
``spin-flipped" density matrix operator
$R=\rho(\sigma_y\otimes\sigma_y)\rho^*(\sigma_y\otimes\sigma_y)$ and
$\sigma_y$ is the Pauli Y matrix, i.e., $\sigma_y= \left(
\begin{array}{cc}
  0 & -i \\
  i & 0
  \end{array}
  \right)$.
Particularly, for a density matrix of the form

\begin{eqnarray}
 \rho=\left(\begin{array}{cccc}
  a & 0 & 0 & 0\\
  0 & b & z & 0  \\
  0 & z^* & c & 0  \\
 0& 0 & 0 & d
        \end{array}
  \right),
\end{eqnarray}
the concurrence is
\begin{eqnarray}
C=2\max\{0, |z|-\sqrt{ad}\}.
\end{eqnarray}

Combing the above equation with the reduced density matrix, we find
that the concurrence of two atoms is
\begin{eqnarray}
C_{a_1a_2}(t)=2|f_1(t)|^2\max\{0, |\alpha \beta|-|\beta f_2(t)|^2\},
\end{eqnarray}
and the concurrence of two cavities is
\begin{eqnarray}
C_{c_1c_2}(t)=2|f_2(t)|^2\max\{0, |\alpha \beta|-|\beta f_1(t)|^2\}.
\end{eqnarray}

In Fig.1, the evolution of two-qubit concurrence for different
partitions $C_{a_1a_2}$ (solid line) and $C_{c_1c_2}$ (dotted line)
are plotted with $\alpha=1/\sqrt{10}, \beta=3/\sqrt{10}, \omega=3,
\omega_0=2, g=1$. For simplicity, we sometimes choose the special
case of $\omega:\omega_0:\omega_c=3:2:1$. On the one hand, the
concurrence of two atoms $C_{a_1a_2}$ will disappear within a finite
time during the dynamics evolution(ESD). On the other hand, the
concurrence of two cavities $C_{c_1c_2}$ can appear during the
dynamics evolution(see the dotted line in Fig.1). It is not
difficult to see that the time for which ESD($t_{ESD}$) and
ESB($t_{ESB}$) occur could be adjusted by controlling the frequency
$\omega_c$ and strength $\lambda$ of classical driving fields. In
addition, the amount of entanglement between two cavities can also
be controlled by classical driving fields.

In order to show this more clearly, we plot the two-qubit
concurrence for different partitions $C_{a_1a_2}$ (solid line) and
$C_{c_1c_2}$ (dotted line) with $\alpha=\sqrt{3}/\sqrt{10},
\beta=\sqrt{7}/\sqrt{10}, \omega=3, \omega_0=2, g=1$ in Fig.2.
Comparing Fig.1 and Fig.2, one can see time of ESD($t_{ESD}$) and
ESB($t_{ESB}$) depend on the parameters $\alpha$ and $\beta$. In the
case of $\alpha=1/\sqrt{10}$ and $\beta=3/\sqrt{10}$,
$t_{ESD}<t_{ESB}$, that is, ESB appears after ESD. However, when
$\alpha=\sqrt{3}/\sqrt{10}$ and $\beta=\sqrt{7}/\sqrt{10}$,
$t_{ESD}>t_{ESB}$, that is, ESB appears before ESD. Again, the time
of ESD and ESB and the amount of entanglement between two cavities
can be controlled by adjusting classical driving fields.

We now turn to show the influence of classical driving fields on the
distribution of entanglement in the present system. The bipartite
entanglement of $a_1\otimes a_2$, $c_1\otimes c_2$, $a_1\otimes
c_2$, and $c_1\otimes a_2$ are displayed in Fig.3. It is not
difficult to see that the concurrence $C_{a_1a_2}$, $C_{c_1c_2}$,
$C_{a_1c_2}$, and $C_{c_1a_2}$ are periodic functions of time t. The
periods of them depend on the strength and the frequencies of
classical driving fields. Comparing the right panel and the left
panel of Fig.3, we find that the time of ESB and ESD and the amount
of the entanglement of two qubits can be controlled by classical
driving fields. For example, $t_{ESD}$ and the amount of
$C_{c_1c_2}$(dashed line) of the right panel are larger than that of
the left panel.

\section{CONCLUSIONS}
In summary, we have considered a quantum system consisting of two
noninteracting atoms each locally interacting with its own vacuum
field. The two atoms, which are driven by two classical fields, are
initially prepared in entangled states. We find that classical
driving fields can increase the amount of entanglement of the
two-atom system. It is worth noting that the time of ESB and ESD can
be controlled by the classical driving fields. The approach
presented in the present Letter may have potential applications in
quantum information processing.

\section*{ ACKNOWLEDGEMENTS}
 This project was supported by the National Natural
Science Foundation of China (Grant No.10774131) and the National Key
Project for Fundamental Research of China (Grant No. 2006CB921403).

\bibliographystyle{apsrev}

\newpage
List of captions

FIG.1 The concurrence of two atoms (solid line) and two caviteis
(dotted line) are plotted as a function of t  with
$\alpha=1/\sqrt{10}, \beta=3/\sqrt{10}, \omega=3, \omega_0=2, g=1$.
Right panel: $\omega_c=\lambda=0$. Left panel: $\omega_c=\lambda=1$.

FIG.2 The concurrence of two atoms (solid line) and two caviteis
(dotted line) are plotted as a function of t  with
$\alpha=\sqrt{3}/\sqrt{10}, \beta=\sqrt{7}/\sqrt{10}, \omega=3,
\omega_0=2, g=1$. Right panel: $\omega_c=\lambda=0$. Left panel:
$\omega_c=\lambda=1$.

FIG.3 The concurrence of two qubits for different partitions are
plotted as a function of t with $\alpha=\sqrt{3}/\sqrt{10},
\beta=\sqrt{7}/\sqrt{10}, \omega=3, \omega_0=2, g=1$. Right panel:
$\omega_c=\lambda=0$. Left panel: $\omega_c=\lambda=1$.

\newpage

\begin{figure}
\centering{\scalebox{0.6}[0.8]{\includegraphics{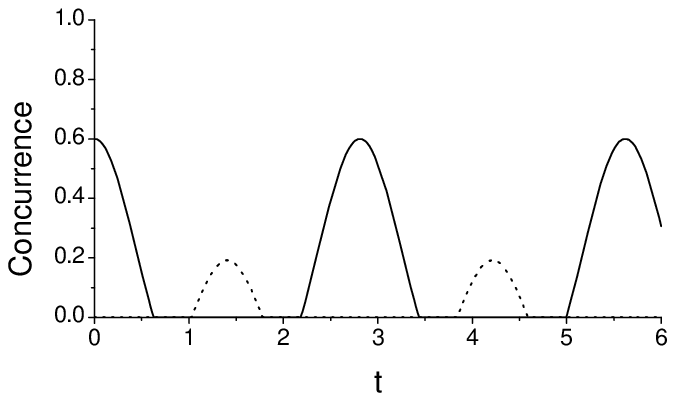}}}
\centering{\scalebox{0.6}[0.8]{\includegraphics{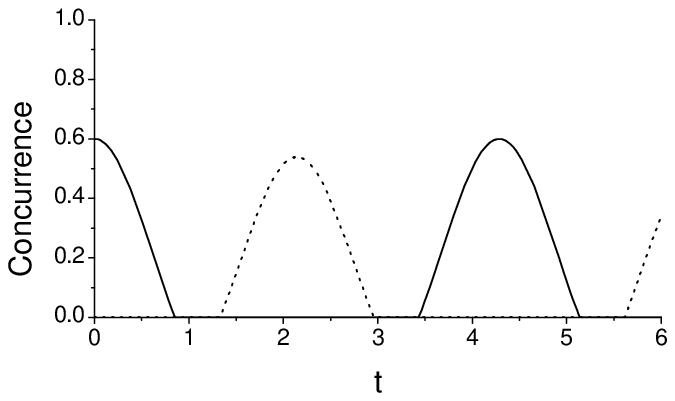}}}
  \caption{The concurrence of two atoms (solid line) and two caviteis (dotted line) are plotted as
a function of t  with $\alpha=1/\sqrt{10}, \beta=3/\sqrt{10},
\omega=3, \omega_0=2, g=1$. Right panel: $\omega_c=\lambda=0$. Left
panel: $\omega_c=\lambda=1$.}
\end{figure}

\begin{figure}
\centering{\scalebox{0.6}[0.8]{\includegraphics{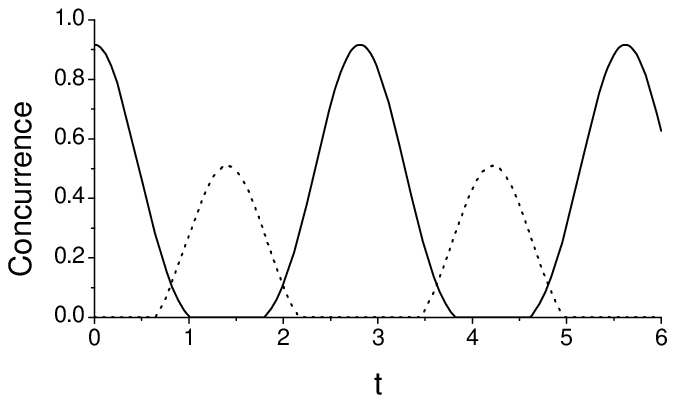}}}
\centering{\scalebox{0.6}[0.8]{\includegraphics{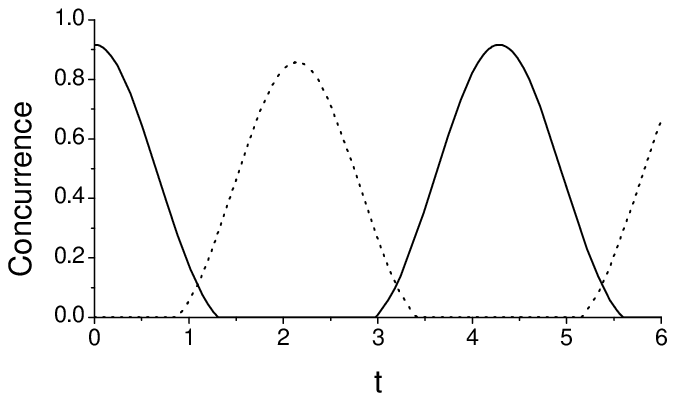}}}
  \caption{The concurrence of two atoms (solid line) and two caviteis (dotted line) are plotted as
a function of t  with $\alpha=\sqrt{3}/\sqrt{10},
\beta=\sqrt{7}/\sqrt{10}, \omega=3, \omega_0=2, g=1$. Right panel:
$\omega_c=\lambda=0$. Left panel: $\omega_c=\lambda=1$.}
\end{figure}

\begin{figure}
\centering{\scalebox{0.6}[0.8]{\includegraphics{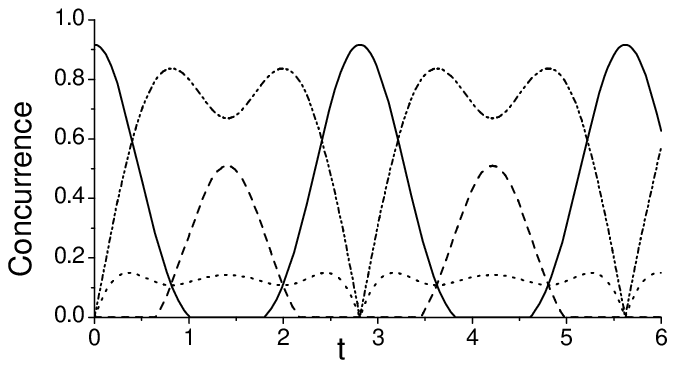}}}
\centering{\scalebox{0.6}[0.8]{\includegraphics{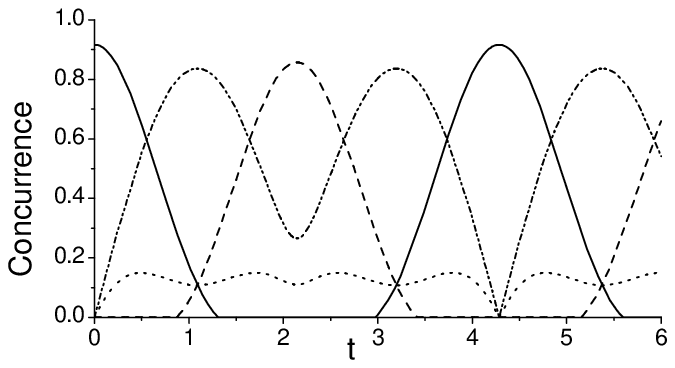}}}
  \caption{The concurrence of two qubits for different partitions are plotted as
a function of t with $\alpha=\sqrt{3}/\sqrt{10},
\beta=\sqrt{7}/\sqrt{10}, \omega=3, \omega_0=2, g=1$. Right panel:
$\omega_c=\lambda=0$. Left panel: $\omega_c=\lambda=1$.}
\end{figure}

\end{document}